\documentclass[preprintnumbers,superscriptaddress,pra,showkeys]{revtex4}
\usepackage{amsfonts}
\usepackage{amssymb}
\usepackage{amsmath}
\usepackage{epsfig}
\usepackage{graphicx}
\usepackage{color}

\input{tcilatex}
\begin{document}

\title{Discrete calculus of variations and Boltzmann distribution without
Stirling's approximation }
\author{Q. H. Liu}
\email{quanhuiliu@gmail.com}
\affiliation{School for Theoretical Physics, School of Physics and Electronics, Hunan
University, Changsha 410082, China}
\date{\today }

\begin{abstract}
A \emph{double extrema form} of the calculus of variations is put forward in
which only the smallest one of the finite differences is physically
meaningful to represent the variational derivatives defined on the discrete
points. The most probable distribution for the Boltzmann system is then
reproduced without the Stirling's approximation, and free from other
theoretical problems.
\end{abstract}

\maketitle

\section{Introduction}

Boltzmann distribution lies in the heart of statistics physics with
applications in mathematics, chemistry and engineering, and it was first
given more than 100 years ago \cite{Boltzmann}. However, seldom feels
comfortable with its derivation with use of the Stirling's approximation 
\cite{61,62,02,1989,2009}. A more puzzling point is that, if one uses more
accurate expression of the Stirling's approximation, or attempts to exact
the derivation of the Boltzmann distribution without invoking the Stirling's
approximation, he always runs into the a surplus $\sim -1/2$ term \cite%
{1989,2009} in the distribution. One can find the relevant derivation in
most standard statistical physics or physical chemistry textbooks, for
instance, \cite{1997,greiner}. The key steps are outlined in the following.

Considering a gaseous system of $N$ noninteracting, indistinguishable
particles confined to a space of volume $V$ and sharing a given energy $E$.
Let $\varepsilon _{i}$ denote the energy of $i$-th level and $\varepsilon
_{1}\prec \varepsilon _{2}\prec \varepsilon _{3}\prec ...$, and $g_{i}$
denote the degeneracy of the level. In a particular situation, we may have $%
n_{1}$ particles in the first level $\varepsilon _{1}$, $n_{2}$ particles in
the second level $\varepsilon _{2}$, and so on, defining a distribution set $%
\left\{ n_{i}\right\} $ \cite{1997}. The number of the distinct microstates
in set $\left\{ n_{i}\right\} $ is then given by,%
\begin{equation}
\Omega \left\{ n_{i}\right\} =\prod_{i}\frac{\left( g_{i}\right) ^{n_{i}}}{%
n_{i}!}.
\end{equation}%
The distribution set $\left\{ n_{i}\right\} $ must conform to two
macroscopic conditions,%
\begin{equation}
\sum_{i}n_{i}=N,\sum_{i}n_{i}\varepsilon _{i}=E.
\end{equation}%
The entropy of the system is $S=S\left( N,V,E\right) =k_{B}\ln \Omega
\left\{ n_{i}^{\ast }\right\} $ where $k_{B}$ is the Boltzmann constant, and 
$\left\{ n_{i}^{\ast }\right\} $ is the most probable distribution,
determined by maximizing the following variational, 
\begin{equation}
f=k_{B}\ln \Omega \left\{ n_{i}\right\} -\alpha \left(
\sum_{i}n_{i}-N\right) -\beta \left( \sum_{i}n_{i}\varepsilon _{i}-E\right) ,
\label{3}
\end{equation}%
where $\alpha $ and $\beta $ are two Lagrangian multipliers. The variational
calculation gives with the discrete indices $i=1,2,3,...$\cite%
{1989,2009,1997,greiner}, 
\begin{equation}
\delta f=\sum_{i}\left( \delta n_{i}\left( \ln g_{i}-\alpha -\beta
\varepsilon _{i}\right) -\delta \ln n_{i}!\right) .  \label{4}
\end{equation}%
The most probable distribution $\left\{ n_{i}^{\ast }\right\} $ satisfies 
\cite{1989,2009,1997,greiner}, 
\begin{equation}
\frac{\delta f}{\delta n_{i}^{\ast }}=\left( \ln g_{i}-\alpha -\beta
\varepsilon _{i}\right) -\frac{\delta \ln n_{i}^{\ast }!}{\delta n_{i}^{\ast
}}=0,\text{ i.e., }\psi (1+n_{i}^{\ast })=\ln g_{i}-\alpha -\beta
\varepsilon _{i},  \label{5}
\end{equation}%
where 
\begin{equation}
\psi (1+x)\equiv \frac{d\ln x!}{dx}
\end{equation}%
is the so-called Digamma function of variable $x$, and $\Gamma (1+x)\equiv x!
$ is the Gamma function.

When $x\gg 1$, utilizing the usual form of the Stirling's approximation\ $%
\ln x!\approx x\ln x-x$, we have, 
\begin{equation}
\psi (1+x)\approx \ln x.  \label{app}
\end{equation}%
In the limit $x\gg 1$, the standard expression for the Boltzmann statistics
is recovered \cite{1989,2009,1997,greiner},%
\begin{equation}
n_{i}^{\ast }\approx g_{i}e^{-\alpha -\beta \varepsilon _{i}}.
\end{equation}%
However, it is not satisfactory. We can use grand ensemble statistics to
prove that the approximately-equal-to symbol "$\approx $" is in fact the
identical-to one "$=$" \cite{greiner}, and a more convenient and detailed
derivation is available from a website lecture note \cite{lectnote}.

The Stirling's approximation\ $\ln x!\approx x\ln x-x$ holds only when $x$
is large. When $x$ is small, a slight deviation from the distribution $%
n_{i}^{\ast }\approx g_{i}e^{-\alpha -\beta \varepsilon _{i}}$ as $%
i\rightarrow \infty $ may cause\ a "long tail" contribution. How about a
more accurate approximation\ of $\ln x!$ is used? \ In the following section
II, we will show the continuous calculus of variations in step (\ref{4}) is
questionable. In section III, with the proper use of the discrete calculus
of variations, the relevant difficulty is removed, and the approximate
relation (\ref{app}) turns out to be an exact result. In section IV, some
further comments on the derivatives in the discrete calculus of variations
are added. The final section V is a brief conclusion.

\section{Appearance of additional $-1/2$ term with the continuous calculus
of variations}

With more accurate expression of $\ln x!$\ as $\ln x!\approx \left(
x+1/2\right) \ln x-x+1/2\ln 2\pi $, we have $d\ln x!/dx\approx \ln
x+1/\left( 2x\right) $, and $\exp (d\ln x!/dx)\approx x+1/2$. Then we come
up against a worse result,   
\begin{equation}
n_{i}^{\ast }\approx -\frac{1}{2}+g_{i}e^{-\alpha -\beta \varepsilon _{i}}.
\label{key}
\end{equation}%
When the energy levels are lower to the lowest, i.e., the number $%
n_{i}^{\ast }$ of particles is large, we have approximately the Boltzmann
distribution $n_{i}^{\ast }\simeq g_{i}e^{-\alpha -\beta \varepsilon _{i}}$.
Meanwhile, the high energy levels are less likely to be populated as $%
g_{i}e^{-\alpha -\beta \varepsilon _{i}}\rightarrow 0$ when $\varepsilon
_{i}\rightarrow \infty $, leading to a "long tail" distribution from (\ref%
{key}), 
\begin{equation}
n_{i}^{\ast }\approx -\frac{1}{2},i\rightarrow \infty .
\end{equation}%
This additional $-1/2$ term presents though the "rigorous treatment" of Eq. (%
\ref{4}) is utilized. To see it, let us consider an asymptotic expression of
the Digamma function $\psi (1+x)$ in limit $x\rightarrow \infty $ which is,
with $O\left( x^{-2}\right) $ denoting a quantity of order $x^{-2}$ \cite%
{psi}, 
\begin{equation}
\psi (1+x)\rightarrow \ln x+\frac{1}{2x}+O\left( x^{-2}\right) ,\text{ }%
x\rightarrow \infty .
\end{equation}%
A surprising property of the Digamma function $\psi (1+x)$ is that we can
numerically verify following equation which holds true for all $x\in \left[
0,\infty \right) $,%
\begin{equation}
e^{\psi (1+x)}\approx x+1/2.  \label{digamma}
\end{equation}%
The largest deviation from $1/2$ occurs at the ending point $x=0$ and $%
e^{\psi (1)}=e^{-\gamma }\approx 0.561$, with $\gamma \simeq 0.577$ is the
Euler--Mascheroni constant. As $x$ increases from zero, the difference $%
e^{\psi (1+x)}-x$ monotonically converges to $1/2$; and at $x=10$ and $10^{2}
$, it gives $0.503$ and $0.500$, respectively. Clearly, Eq. (\ref{key})
turns out to be approximately valid \emph{in whole range} of $n_{i}^{\ast
}\in \lbrack 0,\infty )$. When $x$ is large, $x+1/2\approx x$; however, when 
$x=0$, $e^{\psi (1+x)}\approx 0.561\approx 1/2$ leads to the spurious "long
tail" distribution again. 

The additional $-1/2$ term has to be removed, because the energy levels can
never be \emph{negatively} occupied. However, one can not simply discard the 
$-1/2$ term in the limit $i\rightarrow \infty $, because this term is
associated with a divergence as the number of particles is summed up over
all energy levels, 
\begin{equation}
\sum_{i}^{\infty }n_{i}^{\ast }\approx -\frac{1}{2}\sum_{i=1}^{\infty
}1+\sum_{i=1}^{\infty }g_{i}e^{-\alpha -\beta \varepsilon _{i}},\text{ and }%
\sum_{i=1}^{\infty }1\rightarrow \infty .
\end{equation}%
This situation bears a resemblance to the famous divergence in quantum
electrodynamics. Some feels comfortable with it but other are strongly
against it. Remembering that the lattice quantum fields are helpful to
eliminate the undesired divergences, we are confident that the similar and
proper treatment of the discrete calculus can be used to get rid of the
difficulty. To note that the discrete calculus of variations is a
well-established discipline in mathematics with much wider applications, and
one of the fundamental principles is clear for the derivatives should be
replaced by finite differences \cite{1968,guo,2006,2014}.  

\section{Removal of additional $-1/2$ term with discrete calculus of
variations}

To note that the change of the number of the particle should be integers.
The application of the variational calculation to the Boltzmann system in
Eq. (\ref{4}) must be with care. Since the number of particle is discrete,
we must thus define the Digamma function on discrete lattices $x=0,1,2,3,...$%
. Furthermore, the smallest possible change of the particle can only be $+1$
or $-1$ and can never be an infinitesimal.

The minimum finite difference for $\Delta \ln n!/\Delta n$ may take two
forms. One is the so-called the forward one $\psi ^{+}(1+n)$, and another
the backward one $\psi ^{+}(1+n)$, 
\begin{subequations}
\begin{eqnarray}
\frac{\Delta \ln n!}{\Delta n} &=&\frac{\ln \left( n+1\right) !-\ln \left(
n\right) !}{1}=\psi ^{+}(1+n),\text{ and, }  \label{both1-1} \\
\frac{\Delta \ln x!}{\Delta n} &=&\frac{\ln \left( n\right) !-\ln \left(
n-1\right) !}{1}=\psi ^{-}(1+n).  \label{both1-2}
\end{eqnarray}%
In simple forms, we have, respectively, 
\end{subequations}
\begin{subequations}
\begin{eqnarray}
\psi ^{+}(1+n) &\equiv &\ln (n+1)!-\ln n!=\ln (n+1)\text{, }\left( n\succeq
0\right) ,\text{ and}  \label{both2-1} \\
\psi ^{-}(1+n) &\equiv &\ln n!-\ln (n-1)!=\ln n.  \label{both2-2}
\end{eqnarray}%
Two important and somewhat trivial relations are $\ln (n+1)\succ \ln n$, $%
\left( n\succeq 1\right) $, and when $n\rightarrow 0$, $\ln (n+1)\succ \ln n$
can also be understood in the sense of $1\approx e^{\ln (n+1)}\succ e^{\ln
n}\rightarrow 0$. 

The principles of the calculus of variations indicate that a variation of $%
\ln n!$ is to find \emph{a minimum possible amount }of the changes of $\ln n!
$ when $n$ is changed by a minimum possible value. Thus, $\psi ^{-}(1+n)=$ $%
\ln n$ is singled out. I.e., the approximate result (\ref{app}) happens to
be an exact one. Now we have an exact relation,  
\end{subequations}
\begin{equation}
\text{ }n_{i}^{\ast }=g_{i}e^{-\alpha -\beta \varepsilon _{i}}.  \label{ours}
\end{equation}%
It is right the standard form of the Boltzmann distribution applicable in
the whole meaningful interval $n_{i}^{\ast }\in (0,n_{0}^{\ast })$. 

\section{Why the additional $-1/2$ term occcurs\ from the point of the
discrete calculus of variations}

In above section, we stress that only a special choose of discrete
derivatives among $\Delta \ln n!/\Delta n$ suffices to represent the
discrete variational derivative $\delta \ln n!/\delta n$, and we single out
the minimum one $\psi ^{-}(1+n)=\ln n$. In contrast, if including both $\psi
^{+}(1+n)=\ln (n+1)$ and $\psi ^{-}(1+n)=\ln n$, and using the arithmetic
mean of $\psi ^{+}(1+n)$ and $\psi ^{-}(1+n)$ to present the discrete
variational derivative $\delta \ln n!/\delta n$ instead, we have, 
\begin{equation}
\psi _{\min }(1+n)\equiv \frac{1}{2}\left( \psi ^{+}(1+n)+\psi
^{-}(1+n)\right) =\left( \ln n+\ln (n+1)\right) /2.
\end{equation}%
we run across the unnecessary $-1/2$ term again, for we have, 
\begin{equation}
\exp \left( \psi _{\min }(1+n)\right) =\sqrt{n(n+1)}\approx n+1/2,(n\gg 1).
\end{equation}

Such the discrete\ variational derivative is applicable for the definition
of thermodynamic quantities for the \emph{few-particle systems in the
microcanonical ensemble}. Since in the ensemble both the number of particle
and the energy are fixed, the temperature, for instance, can not be defined
by usual derivative, but the variational derivative instead. Following the
same idea presented in above section, we must choose a minimum possible
value among all possible differences. Recent treatments \cite{few} of the
definition of the temperature and other thermodynamic quantities in the
microcanonical ensemble for small systems demonstrate that we must pick up
the backward difference of $\Delta \ln n!/\Delta n$, based on the completely
different arguments.

\section{Conclusions and discussions}

The calculus of variations is to use a small changes in functions and
functionals, to find maxima and minima of them. Once the functions and
functionals are defined on the continuous intervals, the usual derivatives
are sufficient. However, once they are defined on the discrete points, the
variational derivatives correspond to the many finite differences. We in
fact put forward a \emph{double extrema form} of the calculus of variations
in which only the smallest one of the finite differences is physically
meaningful to represent the variational derivatives. By the \emph{double
extrema form}, we mean that the usual form of the calculus of variations
deals with one extrema only. 

The \emph{double extrema form }of the variational (\ref{4}) means not only $%
\delta f=0$, but also $\delta \ln n!/\delta n=\min \left\{ \Delta \ln
n!/\Delta n\right\} $. The exact form the most probable distribution for the
Boltzmann system is then reproduced without the Stirling's approximation,
and free from other theoretical problems. The \emph{double extrema form} of
the calculus of variations has in fact been used in literature on
statistical mechanics for finite size systems, but based on the physical
arguments.

\begin{acknowledgments}
This work is financially supported by National Natural Science Foundation of
China under Grant No. 11675051.
\end{acknowledgments}


\begin{thebibliography}{99}
\bibitem{Boltzmann} L. Boltzmann, Wissenschaftliche Abhandlungen, Vol. I, F.
Hasen\"{o}hrl (ed.), (Leipzig: Barth, 1909; and New York: Chelsea, 1969
(reissued)).

\bibitem{61} R. W. Hakala, A new derivation of the Boltzmann distribution
law, J. Chem. Educ. \textbf{38}(1), 33 (1961).

\bibitem{62} P. A. H. Wyatt, Elementary statistical mechanics without
Stirling's approximation, J. Chem. Educ.  \textbf{39}(1), 27(1962) .

\bibitem{02} R. B. Shirts, and M. R. Shirts, EDeviations from the Boltzmann
distribution in small microcanonical quantum systems: Two approximate
one-particle energy distributions, J. Chem. Phys. \textbf{117}, 5564(2002).

\bibitem{1989} V. J. Menon and D. C. Agrawal, Method of most probable
distribution: new solutions and results,\ Pramana Journal of Physics \textbf{%
33}, 455(1989).

\bibitem{2009} S. Kakorin, Revision of Boltzmann statistics for a finite
number of particles, Am. J. Phys. \textbf{77}, 48(2009).

\bibitem{1997} R. K. Pathria, Statistical Mechanics, 2nd Ed., (Oxford:
Butterworth-Heinemann, 1997).

\bibitem{greiner} W. Greiner, L. Neise, H. St\"{o}ker, Thermodynamics and
statistical mechanics, (New York: Springer-Verlag, 1995) pp.297.

\bibitem{lectnote} Clare C. Yu, https://ps.uci.edu/\symbol{126}%
cyu/p115A/LectureNotes/Lecture13/lecture13.pdf.

\bibitem{1968} J. A. Cadzow, Discrete calculus of variations, International
Journal of Control \textbf{11}, 393(1970).

\bibitem{guo} Han-Ying Guo, On variations in discrete mechanics and field
theory, Journal of Mathematical Physics \textbf{44}, 5978(2003).

\bibitem{2006} J. N. Reddy, An Introduction to the Finite Element Method
(3rd ed.). (New York: McGraw-Hill, 2006).

\bibitem{2014} K. J. Bathe, Finite Element Procedures, (2nd ed.),
(Watertown: MA. 2014).

\bibitem{psi} M. Abramowitz, and C. A. Stegun, (eds.). \textquotedblleft Psi
(Digamma) Function.\textquotedblright\ in Handbook of Mathematical Functions
with Formulas, Graphs, and Mathematical Tables, 9th printing. (New York:
Dover, 1972) pp. 258-259.

\bibitem{few} E. N. Miranda, Statistical mechanics of few-particle systems:
exact results for two useful models, Eur. J. Phys. \textbf{38}, 065101(2017).
\end{thebibliography}
\end{document}